\documentstyle[prl,aps,twocolumn]{revtex}
\begin{document}

\newcommand{\cvd}{\begin{flushright}$\Box$\end{flushright}}
\newcommand{\ad}{{\rm ad}\;}
\newcommand{\dd}{{\rm d}}
\newcommand{\ee}{{\rm e}}
\newcommand{\tr}{{\rm Tr}\;}
\newcommand{\sgn}{{\rm sgn}\,}
\newcommand{\rif}[1]{(\ref{#1})}
\newcommand{\der}[2]{{\partial #1\over\partial #2}}
\newcommand{\bra}[1]{\left< #1\right|}
\newcommand{\ket}[1]{\left| #1\right>}
\newcommand{\sign}{{\mathrm{sign}}}
\newcommand{\demi}{{1\over 2}}
\newcommand{\inv}[1]{{1\over #1}}
\newcommand{\intS}{\int_\Sigma}
\newcommand{\intM}{\int_{\cal M}}
\newcommand{\intdM}{\int_{\partial\cal M}}
\newcommand{\dM}{{\partial\cal M}}
\newcommand{\M}{{\cal M}}
\newcommand{\RR}{{\cal R}}
\newcommand{\DD}{{\cal D}}
\newcommand{\HH}{{\cal H}}
\newcommand{\VV}{{\cal V}}
\newcommand{\lie}{{\cal L}}
\newcommand{\G}{{\cal G}}
\newcommand{\p}{\partial}
\newcommand{\w}{\wedge}
\newcommand{\rerg}{r_{\rm erg}}
\newcommand{\ig}{{g^{-1}}}
\newcommand{\ih}{{h^{-1}}}
\newcommand{\At}{{\tilde A}}
\hbadness=10000
\newcommand{\lp}{\left(}
\newcommand{\rp}{\right)}
\def\al{\alpha}
\def\adsd{AdS$-2$}
\def\PLB{ Phys. Lett.  }
\def\PLA{ Phys. Lett.  }
\def\CQG{{  Class. Quant. Grav. }}
\def\NCA{{ Nuovo Cimento} }
\def\NPB{{  Nucl. Phys.} }
\def\PLB{{ Phys. Lett.  }}
\def\PRL{ { Phys. Rev. Lett.} }
\def\PRD{{ Phys. Rev.  }}
\def\ZPC{{  Z. Phys. }}
\def\MPLA{{  Mod. Phys. Lett. }}
\def\CMP{{ Commun. Math. Phys. }}
\def\AP{ { Ann. Phys. }}
\def\dx{\int d^2x\ \sqrt{-g}\ }
\def\xp{x^+}\def\xm{x^-}
\def\l{\lambda}
\def\lq{\l^2}
\def\ha{{1\over 2}}
\def\a{\alpha}
\def\b{\beta}
\def\c{\chi}
\def\d{\delta}
\def\f{\phi}
\def\g{\gamma}
\def\i{\iota}
\def\j{\vartheta}
\def\k{\kappa}
\def\o{\omega}
\def\p{\pi}
\def\q{\psi}
\def\r{\rho}
\def\s{\sigma}
\def\t{\tau}
\def\u{\upsilon}
\def\vu{\varphi}
\def\w{\varpi}
\def\y{\eta}
\def\z{\zeta}
\def\D{\Delta}
\def\F{\Phi}
\def\G{\Gamma}
\def\H{\Theta}
\def\L{\Lambda}
\def\O{\Omega}
\def\P{\Pi}
\def\Q{\Psi}
\def\S{\Sigma}
\def\U{\Upsilon}
\def\X{\Xi}
\def\lie{{\cal L}}
\def\de{\partial}
\def\na{\nabla}
\def\per{\times}
\def\inf{\infty}
\def\id{\equiv}
\def\mo{{-1}}
\def\ha{{1\over 2}}
\def\qu{{1\over 4}}
\def\di{{\rm d}}
\def\pro{\propto}
\def\app{\approx}
\def\const{{\rm const}}
\def\ex{{\rm e}}
\def\gmn{g_{\m\n}}
\def\ep{\e_{\m\n}}
\def\ghmn{\hat g_{\m\n}}
\def\mn{{\mu\nu}}  
\def\dix{\int d^2x\ \sqrt{-g}\ }
\def\ds{ds^2=}\def\sg{\sqrt{-g}}
\def\dhx{\int d^2x\ \sqrt{-\hat g}\ }
\def\dex{\int d^2x\ e\ }
\def\xk{\x^{(k)}}
\def\as{asymptotic symmetries }
\def\eom{equations of motion }
\def\pb{Poisson brackets }

\def\ma{{2 M\over \l}}
\def\na{\nabla}
\def\sg{\sqrt{-g}}
\def\hg{\sqrt{-{\hat g}}}
\def\d{\delta}
\def\e{\Phi}
\def\eo{\e_0}
\def\m{\mu}
\def\n{\nu}
\def\h{\hat}
\def\gmn{g_{\m\n}}
\def\hgmn{\hat \gmn}
\def\dx{\int d^2x\ \sqrt{-g}\ }
\def\dhx{\int d^2x\ \sqrt{-\hat g}\ }
\def\ord#1{o\left(#1\right)}
\def\xo{{\x^\perp}}
\def\xp{{\x^\parallel}}
\def\oo{{\o^\perp}}
\def\op{{\o^\parallel}}
\def\Ul{{1\over\l}}
\def\Ht{{\cal H}}
\def\Hx{{\cal H}_x}
\def\Pe{\P_\y}
\def\Ps{\P_\s}
\def\xn{x^{+}}
\def\xm{x^{-}}
\def\r{\rho}
\def\s{\sigma}
\def\x{\chi}
\def\ds{ds^2=}
\def\t{\tau}
\def\st{\scriptstyle}
\def\sst{\scriptscriptstyle}
\def\mco{\multicolumn}
\def\La{\l^2}
\def\i{\infty}
\def\lb{\label}
\def\ads{anti-de Sitter}
\def\adsd{$\rm AdS_{2}$}

\draft
\title{The AdS/CFT Correspondence in Two Dimensions}
\author{Mariano Cadoni and  Paolo Carta}
\address{ Universit\`a degli Studi di Cagliari, Dipartimento di
Fisica and\\INFN, Sezione di Cagliari, Cittadella Universitaria 09042,
Monserrato, Italy.}
\author{Contribution to the Proceedings of the
Euroconference on "Brane New World and Noncommutative Geometry",
Turin,  October 2000}
\maketitle
\begin{abstract}
We review recent progress in understanding the anti-de
Sitter/conformal field theory correspondence in the context of
two-dimensional (2D) dilaton gravity theory.
\end{abstract}

\section{Introduction}

The $d=2$ case \cite{adscft2,CM99} of the correspondence between
gravity on anti-de Sitter (AdS) space and conformal field theory (CFT)
\cite{adscft} is important for several reasons. First of all, the CFT
involved has an infinite dimensional symmetry, so that the theory is
highly constrained.  Analogously to the $d=3$ case \cite{St},
gravitational structures (e.g. black holes) can be investigated using
conformal field theory techniques.  Secondly, \adsd\ appears as
near-horizon geometry of a variety of higher dimensional black holes
not only in string theory but also in the general relativity context
(the Reissner-Nordstrom solution).  Last but not least, being the
simplest case of the correspondence, the AdS$_{2}$/CFT$_{1}$ duality
can be used to test general ideas about the correspondence in
particular and the holographic principle in general. Of particular
conceptual relevance is the fact that it should provide a
correspondence between a field theory (2D gravity) and conformal
mechanics.

Contradicting the general belief that low-dimensional physics is
simpler than the higher-dimensional one, the AdS$_{2}$/CFT$_{1}$ duality 
has many puzzling and controversial features.  These puzzling features are
related with the peculiarities of 2D gravity. In two dimensions there
are at least three (related and unrelated) versions of gravity
theories: dilaton gravity, random surfaces and string theory with 2D
target space (more generally, one can consider string
compactifications on ten-dimensional backgrounds containing \adsd, e.g
\adsd$\times S^{2}\times T^{6}$).  Perhaps the most striking
peculiarity of the 2D case is the topology of the AdS space involved.
Full AdS spacetime in $d=2$ has cylindrical topology, so that its
boundary is not connected, making difficult the identification of the
boundary CFT that should be dual to the gravity theory.  Owing to this
difficulties it is almost impossible to discuss the correspondence in
general. We will focus on 2D dilaton gravity.

\section{Asymptotic symmetries of AdS$_2$ and the conformal group}
Two-dimensional spacetimes with constant negative  curvature,
$R=-2\lambda^{2}$ 
(in the following they will be
referred to as \adsd)
appear as  dynamical solutions of dilatonic gravity  in two
dimensions \cite{CM}. The simplest case is represented by  the
Jackiw-Teitelboim (JT) model \cite{JT},
\begin{equation}\label{JT}
A={1\over2}\int \sg \, d^2x\, \e\left(R+2\lq\right),
\end{equation}
where $\e$ is a scalar
field related to the usual definition of the dilaton $\phi$ by
$\e=\exp(-2\phi)$.
The  model admits black hole solutions, which have  the form \cite{CM}
\begin{eqnarray}\label{e2}
ds^{2}&=&-(\l^2r^2- {2m_{bh}\over\l \eo})dt^2+
(\l^2r^2-{2m_{bh}\over\l \eo})^{-1}dr^2,\nonumber\\
\e&=&\eo \l r.
\end{eqnarray}
An important property of these black hole solutions is that they are
locally equivalent, modulo 2D diffeomorphisms, to the $m_{bh}=0$
vacuum solution.  Moreover, the global feature of the spacetime are
such that the vacuum solution has to be considered as a portion of
full \adsd\ (which is a geodetically complete spacetime). The vacuum
has a null boundary at $r=0$.  In this way we avoid the difficulty
related with the cylindrical topology of \adsd: our reference
spacetime has only one timelike boundary at $r\to\infty$. But we pay
the price of having a singular geodetically incomplete spacetime,
because now it has a ``singularity'' at $r=0$.  

The black hole
solution (\ref{e2}) can be interpreted as a thermodynamic system.  The
black hole mass depends quadratically on both the Hawking temperature
$T$ and the entropy $S_{bh}$ \cite{CM},
\begin{equation}\label{e7}
 m_{bh}^Ë={2\pi^{2}\eo\over \l} \, T^{2}\,,\qquad
 S_{bh}=4 \pi \sqrt{m \eo\over 2 \l}\,.
 \end{equation}
\adsd\ is a maximally symmetric space; it admits, therefore, three
Killing vectors generating the $SO(1,2)\sim SL(2,R)$ group of
isometries.  The asymptotic symmetries of \adsd\ are by definition the
subgroup of the 2D diffeomorphisms group that leaves invariant the
leading term in the $r\to\infty$ asymptotical expansion of the metric
tensor, i.e.~they preserve the large $r$ behavior
\begin{eqnarray}
g_{tt} &=& -\lambda^2r^2 + \gamma_{tt}(t) + {\cal
O}\left(\frac{1}{r^2}\right),
           \nonumber \\
g_{tr} &=& \frac{\gamma_{tr}(t)}{\lambda^3r^3} + 
{\cal O}\left(\frac{1}{r^5}
           \right), \nonumber \\
g_{rr} &=& \frac{1}{\lambda^2r^2} + \frac{\gamma_{rr}(t)}{\lambda^4r^4}
           + {\cal O}\left(\frac{1}{r^6}\right),\nonumber \\
\e &= &\e_0\left(\lambda\rho(t)r + \frac{\gamma_{\e}(t)}{2\lambda
r}\right)
       + {\cal O}\left(\frac{1}{r^3}\right), \label{d3}
\end{eqnarray}
where the fields $\gamma_{\mu\nu},\gamma_{\e},\rho$ parametrize the
first sub-leading terms in the expansion and can be interpreted as
deformations of the boundary of \adsd\ and of the dilaton.

The form of the boundary conditions for the dilaton $\e$ are
determined by requiring consistency with the action of the diff$_{2}$
group.  They require a (asymptotically) nonconstant dilaton, which
breaks the $SL(2,R)$ group of isometries (and in general the
asymptotical symmetries group (ASG) of the metric) of \adsd. This
symmetry breaking is related with the appearance of a central charge
in the related conformal algebra \cite{CM00}.  The asymptotic form
(\ref{d3}) is preserved by infinitesimal diffeomorphisms
$\chi^{\mu}(x,t)$ of the form \cite{CM99},
\begin{eqnarray}\label{d2}
\chi^t &=& \epsilon(t) + \frac{\ddot{\epsilon}(t)}{2\lambda^4r^2} +
           \frac{\alpha^t(t)}{r^4} + {\cal O}\left(\frac{1}{r^5}\right),
           \nonumber \\
\chi^r &=& -r\dot{\epsilon}(t) + \frac{\alpha^r(t)}{r} +
           {\cal O}\left(\frac{1}{r^2}\right).
\end{eqnarray}
Expanding $\epsilon(t)$ in series one finds that  the generators of
the symmetry, $L_{k}$, satisfy a Virasoro algebra
$[L_k,L_l]=(k-l)L_{k+l}$. The ASG of \adsd\ can be, therefore,
identified with the one-dimensional conformal group  acting on
the 
boundary of \adsd. Moreover, the boundary fields  appearing in
Eq. (\ref{d3}) span a representation of  this group. Under the action
of the group  they transform as conformal fields with definite 
conformal dimensions \cite{CM00}.

The Virasoro algebra associated with the ASG can be centrally extended.
The value  of the central charge $C$ of the algebra is particularly
important because enables  one to give a statistical interpretation of
the thermodynamical behavior (\ref{e7}) of the black hole.
For a generic CFT we have the energy-temperature and entropy-mass
relations \cite{cardy},
\begin{equation}\lb{l2}
m_{CFT}={\pi\over  12} \alpha' C T^{2}\,,\qquad
S_{CFT}=2\pi\sqrt{C \,m_{CFT}\over 6}\,.
\end{equation}
The central charge of the algebra can be computed using a canonical
realization of the ASG \cite{CM99,CM00}. The
asymptotic symmetries define charges $J$, which in the canonical formalism
give a realization  of the ASG, trough the Dirac brackets
\begin{equation}\lb{g8}
\{J[\x],J[\o]\}_{DB}=J[[\x,\o]] +C(\x,\o).
\end{equation}
Using the associated deformation algebra one can calculate the central
charge $C(\x,\o)$.
The central charge  was  first calculated in Ref. \cite{CM99}.
The value  $C=24\e_{0}$ was found, which turned out to be wrong by a 
factor of 2 \cite{CM99,CM00} (see also
Ref.  \cite{navarro}).
The puzzle was resolved independently in Refs \cite{CCKM}, \cite{CV}.
In particular in Ref.  \cite{CCKM}, it was shown that because of the
inner boundary of the spacetime at $r=0$, the central charge has an
additional contribution $C_{ent}$ due to the entanglement of states.
$C_{ent}$ can be calculated using
the coordinate transformation, which maps the vacuum $m_{bh}=0$ into the
black hole solution together, together with the anomalous transformation 
law of the energy momentum tensor $T_{tt}$.  One finds $C_{ent}=-12\eo$, 
so
that $C_{tot}= C+C_{ent}=12\eo$.  Inserting this value of the central
charge into the relations (\ref{l2}), one reproduces exactly the
thermodynamical relations (\ref{e7}).
  
\section{The sigma model approach}
Two-dimensional dilaton gravity can be formulated as a nonlinear
sigma model \cite{cav}. For  the JT model  the action reads
\begin{equation}\lb{e8}
A={1\over 2}\int_\Sigma d^2x\, \sg \, {\partial_\mu\phi\partial^\mu M
\over \Phi^{2}-M},
\end{equation}
where $M$ is the mass functional \cite{mass}, which  on the classical 
orbit
is constant and proportional to the ADM mass of the
black hole,   $M=2\eo \l  m_{bh}$.
Expanding near  $\Psi=(- 2\lambda^2\Phi)^{-1}=0 $ one has \cite{CC}
\begin{equation}\lb{e9}
A=\int d^2x\, \sg \, \partial_\mu M\partial^\mu\psi
\left[1+\sum_{k=1}^{+\infty}(2\lambda)^{2k}M^k\psi^{2k}\right].
\end{equation} 
Keeping in mind that $\Phi^{-1}$ is the (coordinate dependent)
coupling constant of the gravitational theory, one easily realizes
that equation (\ref{e9}) is both a weak-coupling  and a near-boundary
(around $r=\infty$) expansion. The leading term describes a free
CFT$_{2}$, which by means of a trivial field redefinition can be cast
in  the form of a  bosonic string with 2D target-space \cite{CC},
\begin{equation}\lb{act}
A_0={1\over 2\pi\alpha'}\int d^2z\,\partial X^\mu\bar\partial
X_\mu.
\end{equation} 
Since AdS$_{2}$ has a timelike boundary the action (\ref{act})
must necessarily describe open strings. On the boundary  we can impose
either 
Dirichlet [$\partial_{a}X^{\mu}(x=0)=0$]  or Neumann
[$n^{a}\partial_{a} X^{\mu}(x=0)=0$, where $n^{a}$ is the normal to the
boundary]  boundary conditions.

Because the weak-coupling expansion (\ref{e9}) is also a near-boundary
expansion,
one expects  the symmetry group of the open string to be related  
with the
ASG of \adsd\ and gravitational asymptotical modes to have an
interpretation in terms of string normal modes.
One can show that the symmetry group of the string can be
obtained from the Killing vectors (\ref{d2}), generating the ASG,
by simply fixing the  higher order terms in the expansion the Killing
vectors (\ref{d2}) (the so-called pure gauge diffeomorphisms).
Moreover the boundary fields $\gamma,\rho$ of Eq. (\ref{d3}) 
have a natural interpretation in
terms of CFT$_{2}$ fields. They transform in the holomorphic sector of
the open string theory as  fields of a given conformal dimension.

We can also write down explicitly the relationship between
gravitational modes $M_{m,n},\Psi_{m,n}$ appearing in the
asymptotical expansion of the fields $M, \Psi$  and string normal modes
$\alpha^\mu_m$. We have \cite{CC}
\begin{eqnarray}\lb{h4}
\alpha^\mu_m&=&
-i \sqrt\pi 2^{-1/2-m}\left( mM_{0,-m}\right),\nonumber\\
\alpha^\mu_m&=&i\sqrt{\pi}2^{-1/2-m}
\left[M_{1,-1-m}\mp\Psi_{1,-1-m}\right],  
\end{eqnarray} 
respectively for  Neumann and Dirichlet boundary conditions.

A crucial difference between the two sets of boundary conditions is
that,  whereas in the Dirichlet case the Virasoro operators $L_{k}$
can be written in terms of local string oscillators, in  the Neumann case
using Eq.(\ref {h4}) we find $L_{k}=0$ identically.
This means that only in the case of Dirichlet boundary conditions the
ASG can be realized using local string oscillators. Neumann boundary
conditions correspond to  the realization of the symmetry given by the 
charges $J$, which has been  described in the previous section.

The main lesson following from the nonlinear sigma model approach can
be summarized as follows. \adsd\ dilaton gravity has two $\Phi
\to\infty$ degeneration limits. The first defines a duality with an
open string with Dirichlet boundary conditions, can be realized using
local string oscillators and describes a AdS$_{2}$/CFT$_{2}$
correspondence.  The second defines a duality with an open string with
Neumann boundary conditions, cannot be realized using local string
oscillators and describes a AdS$_{2}$/CFT$_{1}$ correspondence.  In
the next section we will identify unambiguously the CFT$_{1}$ involved
in the latter correspondence as a conformal mechanics living in the
$r=\infty$ boundary of \adsd.

Also in the Dirichlet case we can use the AdS/CFT duality to describe
the 2D black hole (\ref{d2}) as a CFT object. In particular, we can
reproduce the thermodynamical entropy by counting the degeneracy of
states in the CFT. Computing the central charge of the Virasoro
algebra using its interpretation as Casimir energy, one finds the same
result obtained in the previous section, $C=12\eo$ \cite{CC}, which,
in turn, inserted in Eq. (\ref{l2}) reproduces the Thermodynamical
parameters (\ref{e7}) of the black hole.
   
\section{AdS$_2$ gravity and conformal mechanics}
The previous sections gave us a strong hint about the existence of
a conformal mechanics description of the weak-coupling regime of \adsd\
gravity. Let us now identify unambiguously this conformal mechanics.
Using the boundary expansion (\ref{d3}) in the field equations for
\adsd\ dilaton gravity  following from the action (\ref{JT}) 
and taking the $r\to\infty$ limit, one gets
the dynamics induced by the bulk gravity theory on the boundary \cite{CCKM},
\begin{eqnarray}\label{d5}
\lambda^{-2}\ddot{\rho}- \rho\gamma +\beta&=& 0,\nonumber\\
\dot{\rho}\gamma + \dot{\beta} &=& 0,
\end{eqnarray}
where $\beta={1\over 2}\rho\gamma_{rr}+\gamma_{\e}$ and 
$\gamma=\gamma_{tt}-{1\over 2}\gamma_{rr}$.
These equations of motion define a conformal mechanics. In fact they are
invariant under the diff$_{1}$ group $\delta(t)=\epsilon(t)$ realized
as \cite{CCKM},

\begin{eqnarray}\lb{e8a}
\delta \rho &=& \epsilon\dot{\rho} - \dot{\epsilon}\rho, \nonumber \\
\delta \beta &=& \epsilon\dot{\beta} + \dot{\epsilon}\beta +
                 \frac{\ddot{\epsilon}\dot{\rho}}{\lambda^2}, \nonumber \\
\delta \gamma &=& \epsilon\dot{\gamma} + 2\dot{\epsilon}\gamma -
                  {\stackrel\dots\epsilon\over \lambda^2}.
\end{eqnarray}
The equation of motions (\ref{d5}) describe a  mechanical
system with anholonomic constraints. Alternatively, introducing
the new coordinate $q=\sqrt{\rho/\lambda}$ with  conformal
dimension $-1/2$, one can show that Eq. (\ref{d5}) are equivalent to
the equation \cite{CCKM},
\begin{equation}
\ddot{q} - \frac{g}{q^3} = \frac{\lambda^2}{2}\gamma q, \label{dffeq}
\end{equation}
together with the Hamiltonian constraint
\begin{equation}
\frac{\dot{q}^2}{2} + \frac{g}{2q^2} = -\frac{\lambda^2}{4}\beta.
\label{dffham}
\end{equation}
The equation of motion (\ref{dffeq}) describes the  (IR-regularized)
De-Alfaro-Fubini-Furlan (DFF) conformal mechanics \cite{dff} coupled 
with an
external source $\gamma$. It can be derived from the action \cite{CCKM}
\begin{equation}
I = \int dt \left[\frac 12 \dot{q}^2 - \frac{g}{2q^2} + \frac 14 \lambda^2
    \gamma q^2\right]. \label{1daction}
\end{equation}
This action is invariant  under the  full conformal group
(diff$_{1}$). We can therefore interpret the theory as the CFT$_{1}$
describing the weak-coupling regime of \adsd\ gravity.
In analogy with  CFT$_{2}$ we can write down the associated
energy-momentum tensor \cite{CCKM}
\begin{equation}\label{e4}
T_{tt}=\lambda^{2}(\dot{\rho}\gamma + \dot{\beta}+M)-2\eo \ddot\rho.
\end{equation}
We can think of $\gamma$ as a external source or, alternatively, as
the time-dependent coupling of the harmonic oscillator potential term
$\sim \gamma q^{2}$. Because $\gamma$ is arbitrary and time-dependent,
the mechanical system is nondeterministic and the energy is not
conserved, as it is evident from the Hamiltonian (\ref{dffham}).  We
have here a strong analogy with disordered systems in statistical
mechanics. From the point of view of the 2D gravity theory $\gamma$
encodes the information about the gauge symmetry of the gravity
theory. These facts give a strong hint on the existence of a deep
relationship between gauge symmetries of fields theories and 
nondeterministic dynamical systems.

\section{ A more general correspondence?}
The discussion of the previous sections has been focalized on the 
particular 2D dilaton gravity model (the JT theory) for
which the curvature of the spacetime is everywhere constant.
However, the main results of our approach can be easily generalized
to models admitting solutions that are \adsd\ {\sl only} asymptotically 
\cite{CM99}. The generalization to the other formulations of 2D gravity
mentioned in the introduction (string theory on backgrounds
containing \adsd\ and random surfaces) is more involved.

The \adsd\ gravity theory we have discussed  in this
presentation (or generalization of it)  can be considered as the
``effective theory'' of the two formulation of 2D gravity mentioned 
above.
In the first case as the effective theory originating from  
10D$\to$2D string compactifications. In the second case (random
surfaces) as some sort of  Liouville theory describing the dynamics of
the conformal mode (now identified with the dilaton) 
(See also Ref. \cite{fursaev}).

The main problem of this schema is that our approach works only for
string compactifications that produce \adsd\ endowed with a 
{\sl non constant} dilaton \cite{CA00}. 
We are not able to describe the interesting class of
compactifications characterized by a {\sl constant}  dilaton.
These solutions, whose prototype is the Bertotti-Robinson solution
of general relativity, have a thermodynamical behavior that differs
drastically from that described here. They are characterized by a
$T=0$ degenerate ground state, separated by a mass gap from the
continuous part of the spectrum.
The form of the AdS/CFT correspondence for this class of models is
still an open question.

On the other hand the correspondence we have found between \adsd\
gravity and DFF conformal mechanics coupled with an external source
indicates that the AdS$_{2}$/CFT$_{1}$ correspondence could be a
particular case of a more general correspondence between boundary CFT's
and large N limit of mechanical system.
Progress along this direction has been achieved in Ref.
\cite{CadCa}, and Ref. \cite{CCK}.  In Ref.  \cite{CadCa} it has been
shown that the physical spectrum and physical states of a bosonic
string with 2D target space can be put in correspondence with the large
$N$ limit a mechanical system defined essentially by decoupled
harmonic oscillators.  This mechanical system has a natural
interpretation  in terms of a one-dimensional stochastic
process.  In the second paper \cite{CCK}, it has been shown that the 
large
$N$ limit of Calogero models is equivalent to a CFT.

\end{document}